\documentclass{cs20proc}

\usepackage{multirow}
\usepackage{subfig}

\editors{S. J. Wolk}
\publisher{Zenodo}
\conference{The 20th Cambridge Workshop on Cool Stars, Stellar Systems, and the Sun}
\conferencedate{2018}

\title{Searching for the Origin of Flares in M dwarfs}
\author{Lauren Doyle, $^{1,2}$ Gavin Ramsay, $^{1}$ John G. Doyle$^{1}$}

\affiliation{$^{1}$ Armagh Observatory and Planetarium, College Hill, Armagh, BT61 9DG \\
			 $^{2}$ Mathematics, Physics and electrical Engineering, Northumbria University, Newcastle upon Tyne, NE1 8ST}

\shorttitle{Origin of Flares in M dwarfs}
\shortauthors{L. Doyle et al.}

\abs{We present an overview of K2 short cadence observations for 34 M dwarfs observed in Campaigns 1 - 9 which have spectral types between M0 - L1. All of the stars in our sample showed flares with the most energetic reaching 3$\times10^{34}$ ergs. As previous studies have found, we find rapidly rotating stars tend to show more flares, with evidence for a decline in activity in stars with rotation periods longer than approximately 10 days. We determined the rotational phase of each flare and performed a simple statistical test on our sample to determine whether the phase distribution of the flares is random or if there is a preference for phase. We find, with the exception of one star which is in a known binary system, that none show a preference for the rotational phase of the flares. This is unexpected and all stars in our sample show flares at all rotational phases, suggesting these flares are not all originating from one dominant starspot on the surface of the stars. We outline three scenarios which could explain the lack of a correlation between the number of flares and the stellar rotation phase. In addition we also highlight preliminary observations of DP Cnc, observed in campaigns 16 and 18, and is one of the stars in our extended sample from K2 Campaigns 10 -18 which are still to be examined. }

\begin{document}

\newcommand{\Mdot}{$\dot M$}
\newcommand{\Msun}{$M_{\odot}$}
\newcommand{\Rsun}{$R_{\odot}$}

\maketitle

\section{Introduction}
Solar flares have been studied for over 150 years with the first observation made by Richard Carrington in 1859. In addition, the correlation between sunspots and flaring activity has been studied for decades and it is recognised these two are very closely related. If we apply the same knowledge to the case of stellar flares then we would expect to see these events from active regions which host spots. Despite the comprehensive study of stellar flares over the years, one area which has not been studied in great detail is the rotational phase distribution of flares in M dwarfs. 

M dwarfs are cool, small, main sequence stars with temperatures and radii between 2400 - 3800 K and 0.20\Msun - 0.63\Msun, respectively \citep{gershberg2005solar}. After a spectral type of M4 it is thought these stars become fully convective \citep{hawley2014kepler}. With this in mind they would not posses a tachocline and would therefore, generate their magnetic field using a different mechanism when compared to our Sun. However, despite this, we do see strong flaring activity from these stars, presenting a gap in the understanding of the origin of the flaring activity.  

The lightcurves of M dwarfs show substantial variations in brightness which can be explained by the presence of a large, dominant spot on the surface. This starspot moves in and out of view as the star rotates and presents one way of determining a stars rotation period. So far, this method has provided accurate rotation period measurements for thousands of low mass stars observed using Kepler/K2. 

Stellar flares are a phenomena which have been studied for a century with \cite{bopp1973high} and \cite{gershberg1983characteristics} making some of the first detailed optical observations. Over the years the physics of stellar flares has been studied by many using observations from $\gamma$-rays to radio frequencies.

So far, a small sample of M dwarfs have been used to study the correlations between stellar rotation phase and the number of flares but nothing of significance has been found. These include work by \cite{ramsay2013short}, \cite{hawley2014kepler}, \cite{ramsay2015view} and \cite{lurie2015kepler}, who inspect the phase distribution of flares in a handful of M dwarfs using Kepler/K2 data. Each of these stars shows significant rotational modulation along with flares present at all rotational phases. However, there was no evidence for any correlation between rotational phase and number of flares. 

Launched in 2009, Kepler has revolutionised the field of stellar astrophysics over the 10 years it has been active. During its lifetime it has produced continuous photometric lightcurves for several 100,000 stars changing the way in which we study the activity on stars. However, with the loss of two of its reaction wheels, Kepler was re-purposed in June 2014 as K2 giving new life to the mission and observing fields along the ecliptic for $\sim$ 70 - 80 days. We use K2 short cadence data (SC) of 1 minute exposure to study the flare properties of a sample of M dwarfs in greater detail and aim to address questions regarding the rotational distribution of the flares. 

\section{Our K2 M dwarf Sample} 
To create our final sample we cross-referenced all known M dwarfs sources observed in SC with the SIMBAD catalog. This allowed us to remove any BY Dra stars and any which showed characteristics of a giant (e.g. radii >1\Rsun). Stars which were too faint to show clear detection in the K2 thumb print image were also removed. Our sample consists of 33 M dwarf stars and 1 L dwarf observed in short cadence (SC) with K2 from Campaigns 1-9 (observations made between May 2014 and July 2016). Each target has been observed for

 \begin{table*}[h!t]
\caption{For the stars in our survey which show flaring activity we indicate their observed rotation period; distance; luminosity and the number of flares together with their duration, amplitude and energy.  Gaia Data release 2 parallaxes \citep{gai18} were inverted to calculate the distances of our sample. The errors on the quiescent luminosity represent the combined error from the distance and PanStarrs magnitudes. Three of the stars in our sample did not posses Gaia parallaxes and these distance were taken from the EPIC Catalogue \citep{huber2016vizier} and are marked with an asterisk. This table has been adapted from \cite{doyle2018investigating}.}

   \begin{center}
   \resizebox{\textwidth}{!}{%
   \label{info_table}

	\begin{tabular}{cccccccc}
    \hline 
    
    EPIC ID        & Rotation Period ($P_{rot}$) & Distance                & log($L_{star}$ )        & No. of flares  & Duration Range    & Amplitude Range   & $log(E_{Kp})$    \\ 
                         & days                                    & pc                           &                                 &                       & minutes                 & Flux --                    & ergs                    \\ %
                         &                                            &                                &                                  &                      &                              &                               &                              \\ 
    \hline
    201611969   & > 70                                    &  28.18 $\pm$ 0.04   & 31.679 $\pm$ 0.006  & 15            & 11.8 -- 35.3     & 0.0009 -- 0.0095  & 31.17 -- 32.18  \\ 
    205204563   & 42:                                      &  14.54 $\pm$ 0.03  & 31.266 $\pm$ 0.007  & 5             & 8.83 -- 20.6     & 0.0003 -- 0.0030  & 29.12 -- 31.05  \\ 
    205467732   & 1.321 $\pm$ 0.021             &  28.3*                      & 30.88                       & 221           & 8.83 -- 91.2     & 0.0003 - 1.3694   & 30.60 -- 33.49  \\ 
    206019387   & > 70                                    & 4.676 $\pm$ 0.002  & 30.864 $\pm$ 0.001   & 47            & 10.8 -- 81.4     & 0.0004 -- 0.0151  & 30.01 -- 31.77  \\ 
    206050032   & > 70                                    & 10.88 $\pm$ 0.01     & 30.564 $\pm$ 0.004  & 17           & 10.8 -- 32.4     & 0.0274 -- 0.6937  & 31.25 -- 32.52  \\ 
    206053352   & --                                        &  39.47 $\pm$ 0.31    & 29.261 $\pm$ 0.032    &  8              & 11.8 -- 30.4     & 0.2254 -- 6.7031  & 31.19 -- 32.27  \\ 
    206262336   & 9.6 $\pm$ 1.2                    & 11.21 $\pm$ 0.02      & 30.385 $\pm$ 0.006    & 237            & 8.78 -- 46.1     & 0.0018 -- 1.2462  & 29.45 -- 33.06  \\ 
    210317378   & 24.5:                                  &  34.97 $\pm$ 0.08     & 31.119 $\pm$ 0.009    & 7              & 12.8 -- 21.6     & 0.0069 -- 0.0582  & 31.39 -- 32.28  \\ 
    210434433   & 47:                                     &  30.90 $\pm$ 0.06     & 31.188 $\pm$ 0.008    & 4              & 14.7 -- 24.5     & 0.0041 -- 0.0207  & 31.02 -- 32.11  \\ 
    210460280   & 45:                                     & 36.22 $\pm$ 0.06     &  31.389 $\pm$ 0.006   & 1             & 29.42            & 0.0024            & 31.47           \\ 
    210489654   & > 80                                   & 28.80 $\pm$ 0.12     &  31.011 $\pm$ 0.016   & 24             & 10.8 -- 39.2     & 0.0037 -- 0.0632  & 31.06 -- 32.28  \\ 
    210579749   & 23:                                     & 17.24 $\pm$ 0.01     & 31.661 $\pm$ 0.003    & 4              & 12.8 -- 42.2     & 0.0014 -- 0.0027  & 31.37 -- 31.95  \\ 
    210758829   & 0.4539 $\pm$ 0.0027        & 18.12 $\pm$ 0.03     & 30.330 $\pm$ 0.007    & 197            & 8.78 -- 106      & 0.0023 -- 4.3192  & 30.09 -- 33.46  \\ 
    210764183   & 0.966 $\pm$ 0.011            & 29.31 $\pm$ 0.36      & 28.936 $\pm$ 0.049   & 10             & 8.78 -- 45.9     & 0.2365 -- 5.1922  & 30.62 -- 31.95  \\ 
    210811310   & 33:                                     & 52.45 $\pm$ 0.12      & 31.477 $\pm$ 0.009   & 2              & 10.8             & 0.0013 -- 0.0028  & 30.88 -- 31.16  \\ 
    210894955   & 0.726  $\pm$ 0.006           & 28.83 $\pm$ 0.11      & 29.763 $\pm$ 0.015   & 18            & 10.8 -- 30.4     & 0.0455 -- 2.3884  & 30.65 -- 32.58  \\ 
    211046195   & 0.2185 $\pm$ 0.0006         & 51.21 $\pm$ 0.40     & 29.988 $\pm$ 0.032   & 16             & 0.98 -- 2888     & 0.0437 -- 4.5665  & 31.15 -- 34.57  \\ 
    211069418   & 0.8177 $\pm$ 0.0086        & 122.9 $\pm$ 9.9       & 31.355 $\pm$ 0.321    & 104          & 8.78 -- 63.8     & 0.0101 -- 0.4138  & 31.65 -- 33.81  \\ 
    211077349   & 0.6992 $\pm$ 0.0061         & 136.3 $\pm$ 1.3      & 31.306 $\pm$ 0.038    & 64             & 8.78 -- 75.6     & 0.0142 -- 3.8455  & 31.69 -- 34.77  \\ 
    211082433   & 0.3447 $\pm$ 0.0015        & 79.24 $\pm$ 1.09     & 31.131 $\pm$ 0.055    & 123           & 8.78 -- 71.6     & 0.0097 -- 0.4683  & 31.17 -- 33.51  \\ 
    211112686   & 0.7639 $\pm$ 0.0073        & 133.9 $\pm$ 0.9        & 31.623 $\pm$ 0.028   & 53             & 8.78 -- 94.0     & 0.0103 -- 0.4311  & 31.99 -- 34.24  \\ 
    211117230   & 0.398 $\pm$ 0.002             & 131.9 $\pm$ 0.9       & 32.016 $\pm$ 0.027   & 38             & 8.83 -- 76.5     & 0.0045 -- 0.2136  & 32.08 -- 34.28  \\ 
    211642294   & 52:                                     & 17.82 $\pm$ 0.02      & 31.307 $\pm$ 0.004  & 5              & 20.6 -- 103      & 0.0023 -- 0.0196  & 31.15 -- 34.42  \\ 
    211945363   &  > 70                                  & 37.7*                       & 31.41                       & 8               & 10.8 -- 31.2     & 0.0025 -- 0.0231  & 31.19 -- 31.98  \\ 
    211970427   & 4.38 $\pm$ 0.24                & 189.3 $\pm$ 5.4        & 31.721 $\pm$ 0.113  & 51            & 8.78 -- 64.8     & 0.0101 -- 0.3572  & 32.04 -- 34.09  \\ 
    212009427   & 1.556 $\pm$ 0.029           & 184.2 $\pm$ 1.2        & 32.083 $\pm$ 0.026  & 75             & 8.78 -- 76.5     & 0.0068 -- 0.0967  & 32.14 -- 34.29  \\ 
    212029094   & 20.22 $\pm$ 5.03             & 185.3 $\pm$ 2.1        & 31.513 $\pm$ 0.045   & 3             & 14.7 -- 22.5     & 0.0019 -- 0.0087  & 31.19 -- 31.66  \\ 
    212518629   & 80:                                    & 21.98 $\pm$ 0.58      & 30.832 $\pm$ 0.105   & 1             & 27.46            & 0.0106            & 31.46           \\ 
    212776174   & 18.35 $\pm$ 5.03             & 24.21 $\pm$ 0.03      &32.022 $\pm$ 0.005    & 7             & 10.8 -- 20.6     & 0.0007 -- 0.0038  & 31.21 -- 31.99  \\ 
    212826600   & --                                      & 109*                         & 30.49                        & 7             & 12.8 -- 41.2     & 0.0015 -- 0.8707  & 32.15 -- 33.31  \\ 
    228162462   & 3.9 $\pm$ 0.2                  & 7.962 $\pm$ 0.004     & 30.156 $\pm$ 0.002   & 424        & 8.78 -- 165      & 0.0015 -- 0.8707  & 29.32 -- 32.74  \\ 
    \hline
    \end{tabular}
    }
	\vspace{0.1cm}\scriptsize{\begin{flushleft}
\textbf{Notes.} For a handful of sources the apparent modulation period is longer than the observation length meaning only an lower limit to the rotation period could be determined. For stars with no evidence for a modulation no rotation period could be determined.
 \end{flushleft}}
 \end{center}
\end{table*}

$\sim$ 70 -80 days producing continuous lightcurves. For this study short cadence data is of particular importance as it provides a complete overview of the flaring activity on these stars. The full stellar properties of our sample can be found in \cite{doyle2018investigating}. 

Due to the positioning of K2 there is a significant level of data preparation needed in the raw K2 lightcurves. For our sample we used the corrected K2 data using the EPIC Variability Extraction and removal for Exoplanet Science Targets ({\tt EVEREST}) pipeline \citep{luger16} for all but one star in our sample. The SC data of GJ 1224 observed in Campaign 9 was obtained from Andrew Vanderberg. Full details of the data preparation and pipeline ca be found in \cite{doyle2018investigating}. 

\section{Determining the Rotation Period}
To determine the rotation period of each star in our sample we initially use a Lomb-Scargle (LS) periodogram, defining phase zero, $\phi$ = 0.0, as the flux minimum of the rotational phase first determined by eye. Stars with rotation modulation longer than 10 days had rotation periods derived using the LS periodogram where phase zero was obtained by folding the lightcurve. Similarly, for stars with rotation modulation less than 10 days we are able to refine the rotation period, $P_{rot}$, and phase zero, $t_{0}$, by phasing and folding sections of the lightcurve taken from the start, middle and end. The iterative procedure allows $P_{rot}$ and $t_0$ to be determined more accurately as these values produce mean folded lightcurves which are used during the remainder of the analysis. Uncertainties are also obtained on each of the rotation periods through the Full Width at Half Max (FWHM) of the corresponding peak on the power spectrum. Table \ref{info_table} shows the rotation periods determined from our sample which range from 0.21 days to greater than 80 days \citep{doyle2018investigating}. It is important to note six stars in our sample possessed incomplete rotational modulation in their lightcurve and so, it is only possible to say they have rotation periods which are greater than the observation length of 70 days.

\section{Identifying Flares}
In order to identify flares within the lightcurve we use a suite of IDL programs, Flares By Eye ({\tt FBEYE}), created by J.R.A. Davenport \citep{davenport2014kepler}. {\tt FBEYE} scans the lightcurve and flags up any points which are over the 2.5$\sigma$ threshold and when there are two or more consecutive points these are identified as potential flares. From this, {\tt FBEYE}, outputs a list of flares which can be read through an interactive GUI to manually classify and remove anything which does not conform to a flare profile (i.e. sharp rise and exponential decay). The final result is a complete list of flares and properties including the peak time, start time, end time, flux peak and equivalent duration \citep{doyle2018investigating}. We now show the number of flares and the range in duration and amplitude, along with the rotation period for each star in Table \ref{info_table}. Three of our sources did not display any flaring activity and were omitted from Table \ref{info_table} and any further analysis, reducing our sample size to 31.

\begin{figure}[h]
	\centering
	\includegraphics[width = 0.5\textwidth]{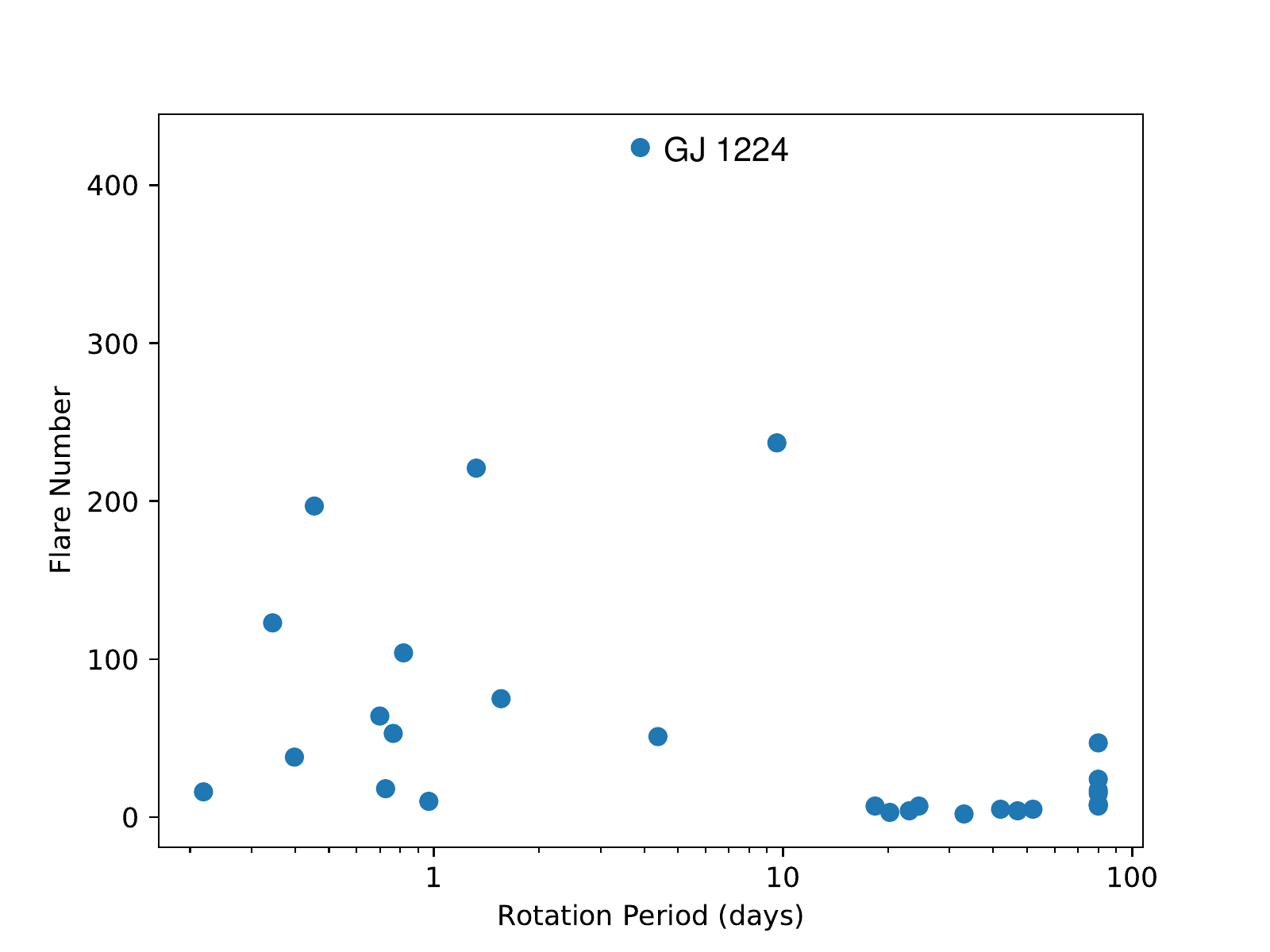}
	\caption{The number of flares as a function of the rotation period for all stars showing flaring activity. Another studies have found, stars with rotation periods longer than $\sim$ 10 days are less active. The most active star is GJ 1224, $P_{rot}$ = 3.9 days. There are six sources to the very right hand side which represent the stars with rotation periods > 70 days. This plot was taken from \cite{doyle2018investigating}. }
	\label{rotation}
\end{figure}

It expected to find increased flaring activity on stars with shorter rotation periods. We find in our sample (Figure \ref{rotation}) there is a decline in the number of flares after a rotation period of 10 days, which is consistent with previous work by \cite{stelzer2016rotation}. However, in order to make complete conclusions about this sample we would have to consider the ages of the M dwarfs, as despite the activity depending on rotation, the rotation in turn depends on age. 

\section{The Flare Energies}
Next we want to calculate the energies of the flares to use in further analysis. To do this the quiescent luminosity of each star, $L_{*}$ must be calculated in the Kepler bandpass. Using PanStarrs magnitudes g, r, i and z (see \cite{doyle2018investigating}) template spectral energy distributions were constructed for each star. By fitting a polynomial to the PanStarrs data and convolving with the Kepler response function, similar to \cite{kowalski2013time}, the flux in the Kepler bandpass can be obtained. The quiescent luminosity is then the flux multiplied by $4\pi d^2$, where distances ($d$) to the stars were obtained through Gaia parallaxes. Full details of this process can be found in \cite{doyle2018investigating} and the distances and luminosities are displayed in Table \ref{info_table}. 

Finally, the energy of the flares (see Table \ref{info_table}) is expressed as the multiplication of the luminosity of the star, $L_{*}$, in erg/s and the equivalent duration, $t$, in seconds. The equivalent duration is defined as the area under the flare lightcurve in units of seconds \citep{gershberg1972some} and was obtained through the FBEYE suite of programs through a Trapezoidal summation.

\begin{figure}
	\centering
	\includegraphics[width = 0.5\textwidth]{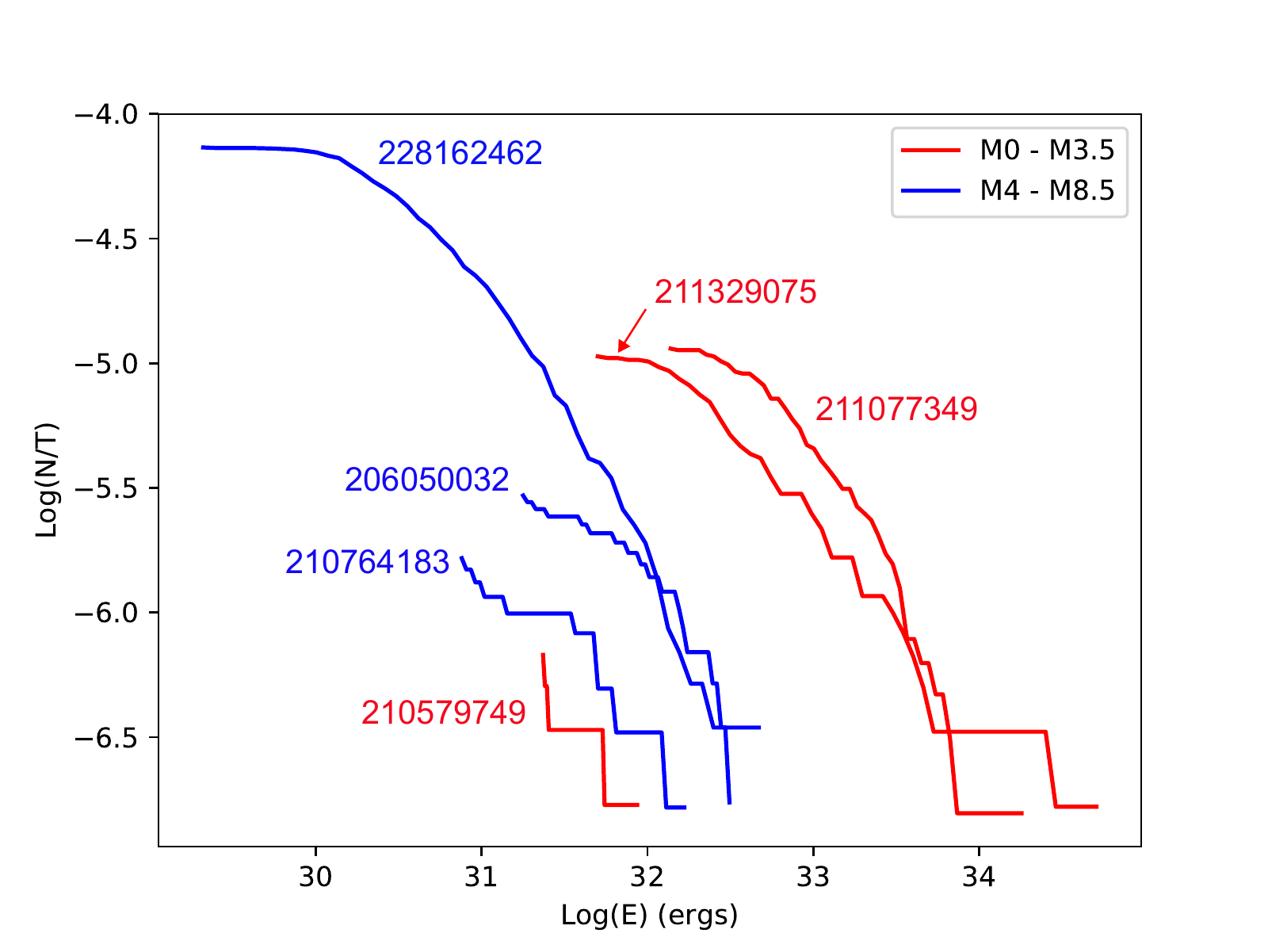}
	\caption{The logarithm of cumulative flare frequency in minutes against the logarithm of flare energies for a small selection of the M dwarf sample with various spectral types, rotations and flare number. The labels represent the EPIC ID for each star shown in this plot and the red and blue are M0 - M3.5 and M4 - M8.5, respectively. }
	\label{frequency}
\end{figure}

Figure \ref{frequency} displays the cumulative frequency distribution of a handful of stars ranging in spectral type and rotation period. Overall, this is telling us the flare rate does not strictly depend on spectral type and rotation does play a role. However, there does seem to be an increase in flaring activity in M4-M8.5 stars but this could be a result of the fully convective regime. A flattening trend is noticed in Figure \ref{frequency} at various energies depending on the star. This has been mentioned in previous work suggesting higher and lower energy flares follow a different power law slope. Therefore, in SC data it is in fact due to a detection limit but is in fact a real feature.

\section{The Rotational Phase}
\begin{figure*}[h!]
	\centering
	\includegraphics[width = 1.0\textwidth]{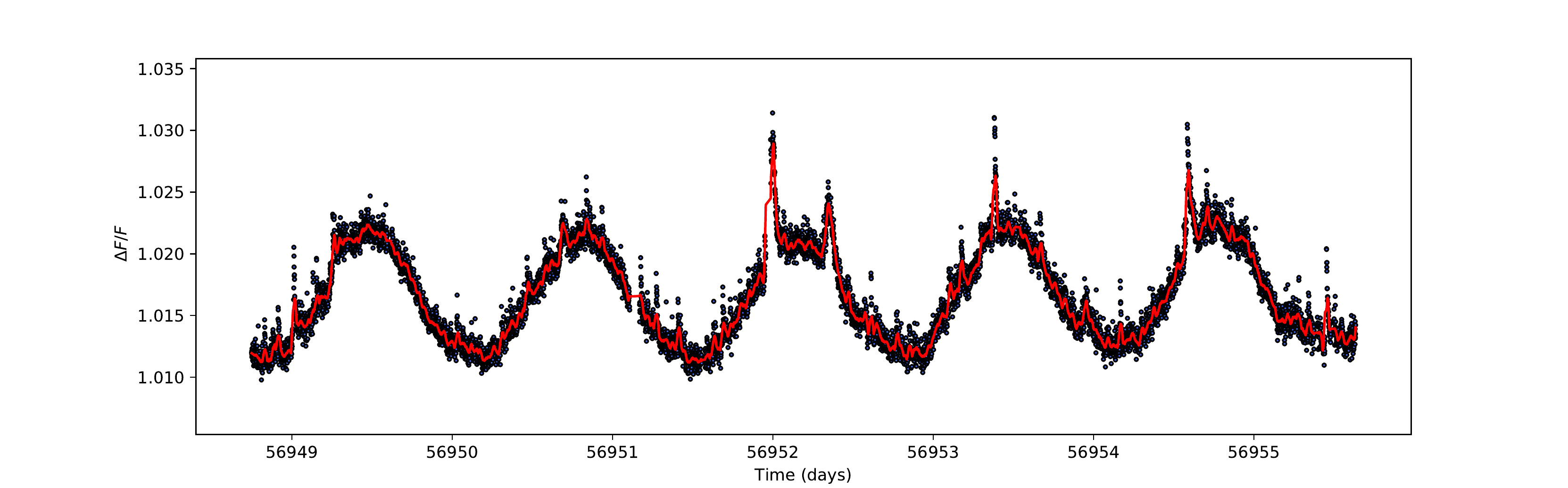}
	\caption{This is a section of the K2 lightcurve for GJ 3954 (EPIC 205467732) from Campaign 2 covering $\sim$ 7 days, which has a rotation period, $P_{rot}$, of 1.321 days. The black points represent the K2 data points and the red line is the Savitzky-Golay filtered, smoothed data. here we can see there are large flares which are present during the maximum peak of rotation when the starspot is least visible. In addition there are smaller flares present throughout the lightcurve at all rotational phases. }
	\label{sec_lightcurve}
\end{figure*}

Stars with large, dominant starspots display periodic changes in their brightness as the rotate, see Figure \ref{sec_lightcurve}. If we assume the same physical processes occur in stellar flares as solar flares, then you would expect the flares to originate from active regions which commonly host spots. Therefore, there should be clear correlations with the rotational phase when the spot is most visible and the number of flares.  

\begin{figure*}[h!]
	\centering
	\subfloat[]{\includegraphics[width = 0.5\textwidth]{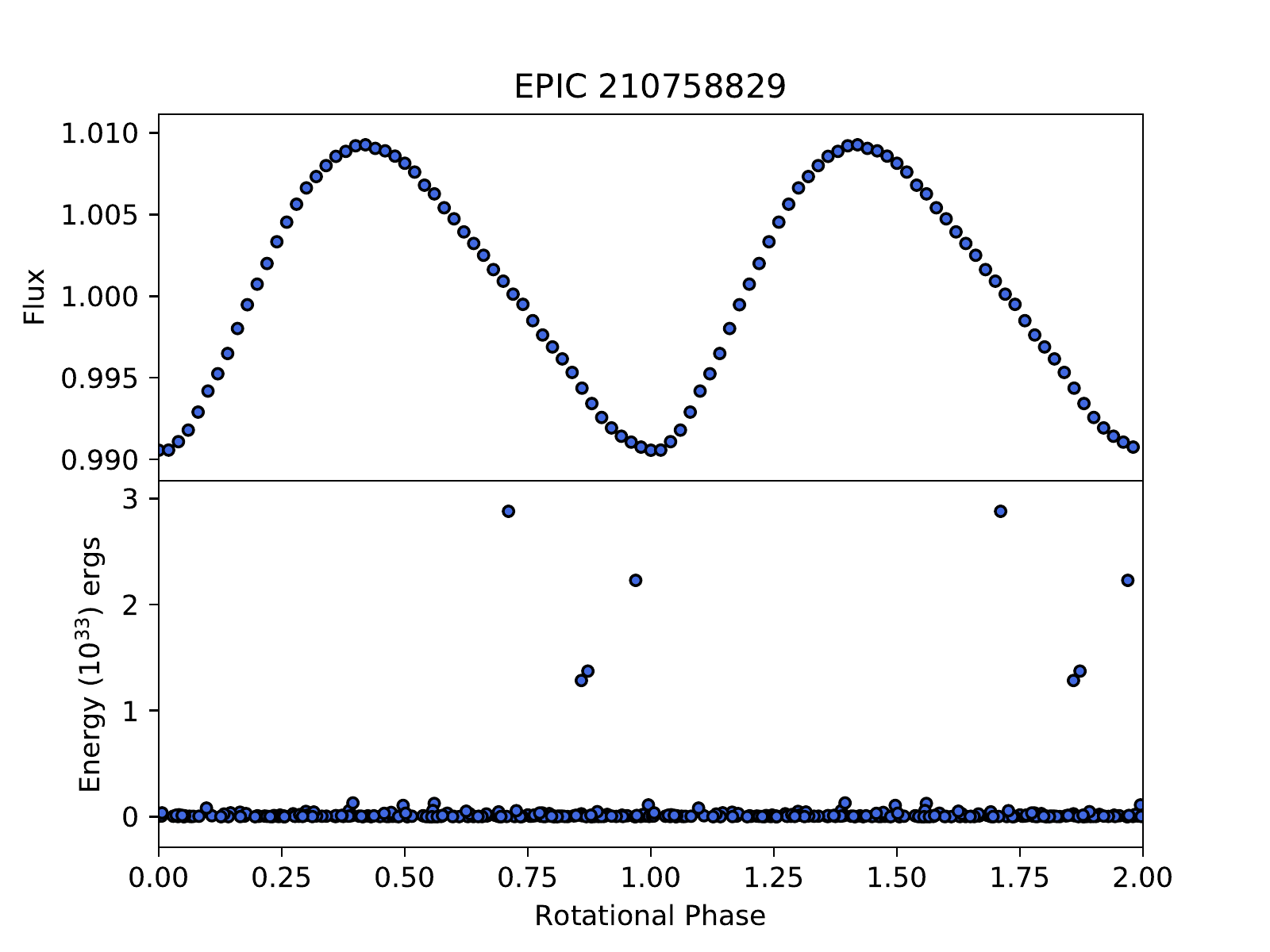}}
	\subfloat[]{\includegraphics[width = 0.5\textwidth]{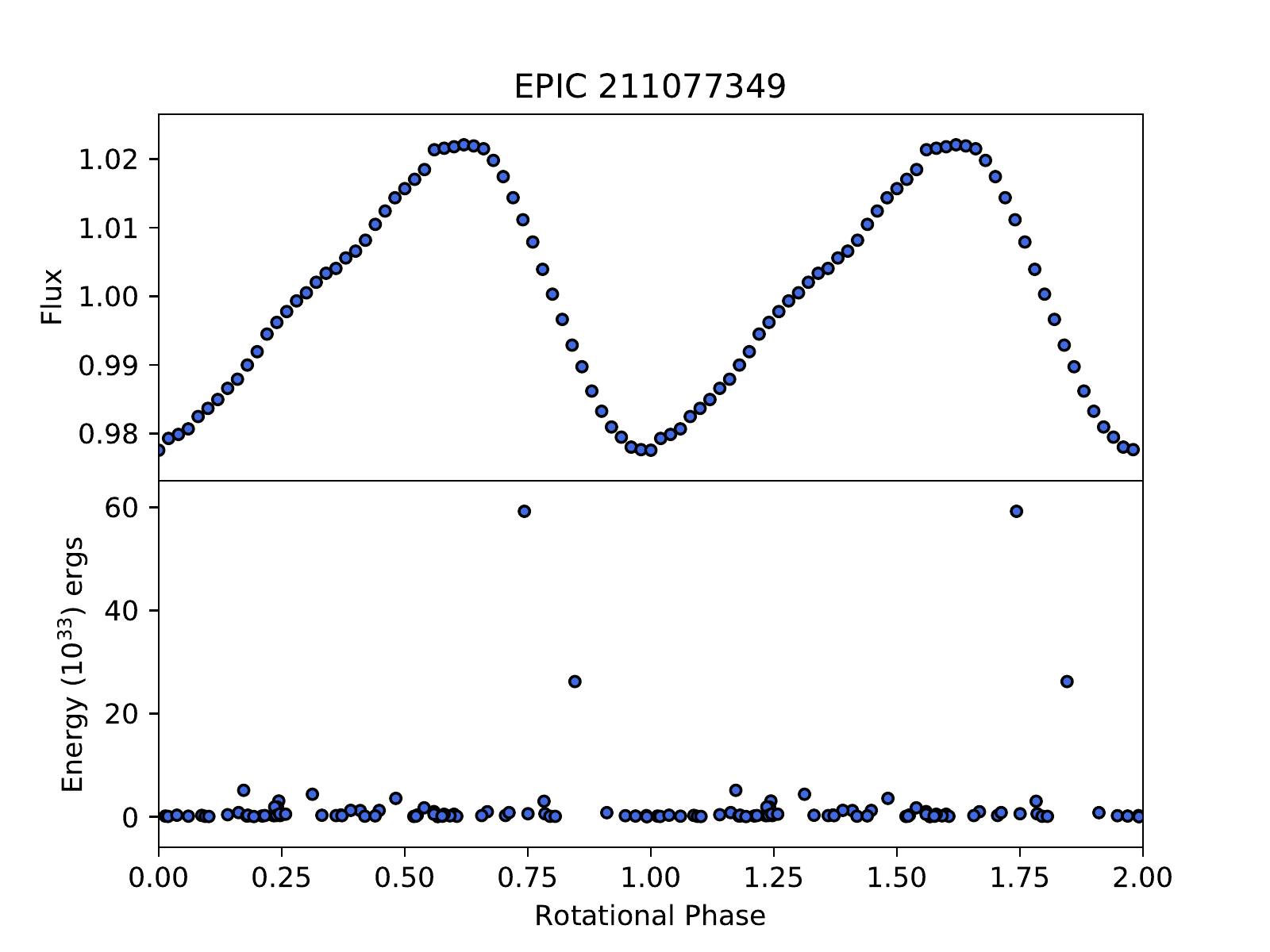}}
	\caption{The top panel of each figure shows the lightcurves of the star phased and binned on the rotation period for 50 bins. The bottom panel shows the phase of the flares with their energy. The data in both plots has been plotted twice so they cover rotation phase 0.0 - 2.0 where 1.0 - 2.0 is simply a repeat. }
	\label{rot_phase}
\end{figure*}

To investigate the behaviour of the flares in relation to the rotational phase we only select stars with rotation periods shorter than the observation length. As a result of this, eight of our sources are omitted from further analysis. We phase fold and bin the lightcurves of our remaining sample using rotation periods and phase zeros determined previously. This produces plots similar to those shown in Figure \ref{rot_phase} for each star. It is noticed that flares are present at all rotational phases for all stars. Additionally, higher energy flares are present at rotation maximum when the starspot is least visible (i.e. behind the disk). The majority of our sample show clear sinusoidal modulation caused by the presence of one dominant, large starspot. 

To test whether the phase distribution of the flares is random we used a simple $\chi^2$ statistical test. The flares were split into low and high energy with the cut-off determined as the median energy of flares for each star. Similarly, the rotational phase was split into 10 bins and $\chi^2$ computed for low, high and all energy categories. Table \ref{chi_squared_results} shows the results for this test \citep{doyle2018investigating}. Out of the flare categories none of the stars show a preference for any rotational phase at a 2$\sigma$ confidence level. This is surprising as it indicates many of the flares do not originate from the large, dominant spot. 

\begin{table}[t]
   \caption{We show the $\chi^{2}$ value for whether each rotation phase bin (split into ten) had flares which were randomly distributed by phase. We split the flares into low and high energy where the cut off is determined by the median energy of all the flares for each star. None of the stars in our sample show a preference for flares at a certain rotational phase. Taken from \cite{doyle2018investigating}. }
   \begin{center}
     \begin{tabular}{ccccc}
      \hline 
     
     \multirow{2}{*}{EPIC} & median energy & \multicolumn{3}{c}{Reduced Chi-Squared} \\
        \cline{3-5}
         & (erg) & low & high & all \\
         \hline
         205204563 & 9.5 $\times 10^{30}$ & 0.78 & 2.00 & 1.00\\
         205467732 & 2.3 $\times 10^{31}$ & 1.17 & 1.17 & 0.55\\
         206262336 & 6.5 $\times 10^{30}$ & 0.92 & 1.13 & 0.61\\
         210317378 & 6.2 $\times 10^{31}$ & 0.78 & 1.22 & 0.97\\
         210434433 & 4.9 $\times 10^{31}$ & 0.89 & 0.89 & 1.22\\
         210579749 & 3.9 $\times 10^{31}$ & 0.89 & 0.89 & 0.67\\
         210758829 & 6.7 $\times 10^{30}$ & 1.14 & 0.30 & 0.62\\
         210764183 & 2.3 $\times 10^{31}$ & 1.44 & 1.00 & 1.11\\
         210811310 & 1.1 $\times 10^{31}$ & 1.00 & 1.00 & 2.00\\
         210894955 & 1.9 $\times 10^{31}$ & 1.10 & 1.35 & 0.47\\
         211046195 & 3.7 $\times 10^{32}$ & 0.33 & 0.78 & 0.78\\
         211069418 & 1.8 $\times 10^{32}$ & 0.93 & 1.33 & 0.65\\
         211077349 & 3.1 $\times 10^{32}$ & 0.82 & 0.82 & 1.15\\
         211082433 & 1.2 $\times 10^{32}$ & 1.11 & 0.39 & 0.72\\
         211112686 & 4.8 $\times 10^{32}$ & 0.58 & 1.21 & 1.18\\
         211117230 & 1.9 $\times 10^{33}$ & 0.87 & 0.99 & 0.69\\
         211642294 & 8.6 $\times 10^{31}$ & 0.78 & 0.89 & 1.00\\
         211970427 & 5.5 $\times 10^{32}$ & 0.82 & 0.82 & 1.24\\
         212009427 & 7.8 $\times 10^{32}$ & 0.54 & 0.81 & 0.33\\
         212029094 & 2.5 $\times 10^{31}$ & 1.00 & 0.89 & 1.52\\
         212776174 & 4.3 $\times 10^{31}$ & 1.22 & 0.78 & 0.97\\
         228162462 & 4.6 $\times 10^{30}$ & 0.83 & 0.48 & 0.94\\
         \hline
      \end{tabular}    
    \end{center}
    \label{chi_squared_results}
\end{table}

\section{Conclusions}
Through the analysis of K2 short cadence data of 34 M dwarfs we have found a surprising but interesting result. We find no correlation between the rotational phase and number of flares. However, all of the stars show significant modulation caused by one dominant starspot and this result suggests these flares do not originate there. To explain this phenomenon we propose three possible explanations. 

\begin{enumerate}
\item The flares could be a result of magnetic interaction between a second star in a binary system. Interactions between the M dwarf and its companion could increase magnetic activity producing flares in locations not hosted by the dominant spot. 
\item Similarly, there is the potential for magnetic interactions between the M dwarfs and orbiting planets. Depending on the number of planets and their properties, the star-planet system could cause increased magnetic activity. 
\item Lastly, there is the possibility of polar spots present on the M dwarf. This would depend on the inclination ($vsini$) of the star. However, if the stars axis was inclined towards the observer, the polar spot can be seen at all phases, interacting with active and quiet regions as the star rotates. This would produce flaring activity at all phases as observed in the data. 
\end{enumerate}

For a more comprehensive explanation on these three scenarios and for full details on all work in this paper refer to \cite{doyle2018investigating}. To test these three scenarios further we would need to investigate the inclination and potential star-planet or binary stars in greater detail. Furthermore, there are the remaining Kepler/K2 Campaigns plus TESS which will allow for a larger sample size and more complete evaluation across the M dwarf spectral range. 

\section{Future Work}
\begin{figure}
	\centering
	\includegraphics[width = 0.5\textwidth]{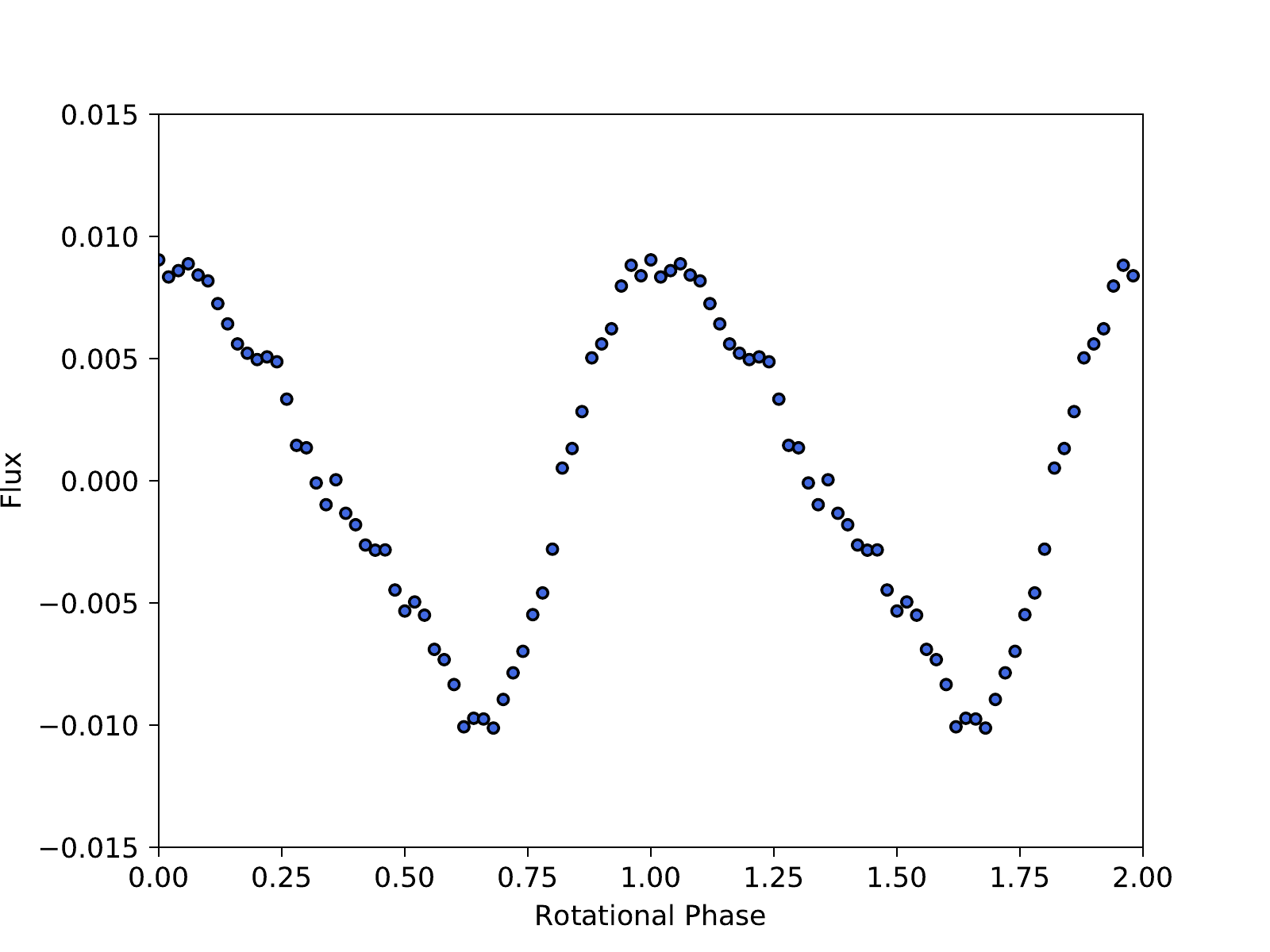}
	\caption{This is the phase folded and binned K2 lightcurve for M dwarf DP Cnc (EPIC 211817361) folded on a rotation period, $P_{rot}$, of 0.596 days. DP Cnc has a spectral type of M3.5 and was observed in Campaigns 16 and 18. Phase 1.0 - 2.0 is simply  repeat of 0.0 - 1.0 and is shown for completeness. }
	\label{sample_lightcurve}
\end{figure}

\begin{figure*}[t]
	\centering
	\includegraphics[width = 1.0\textwidth]{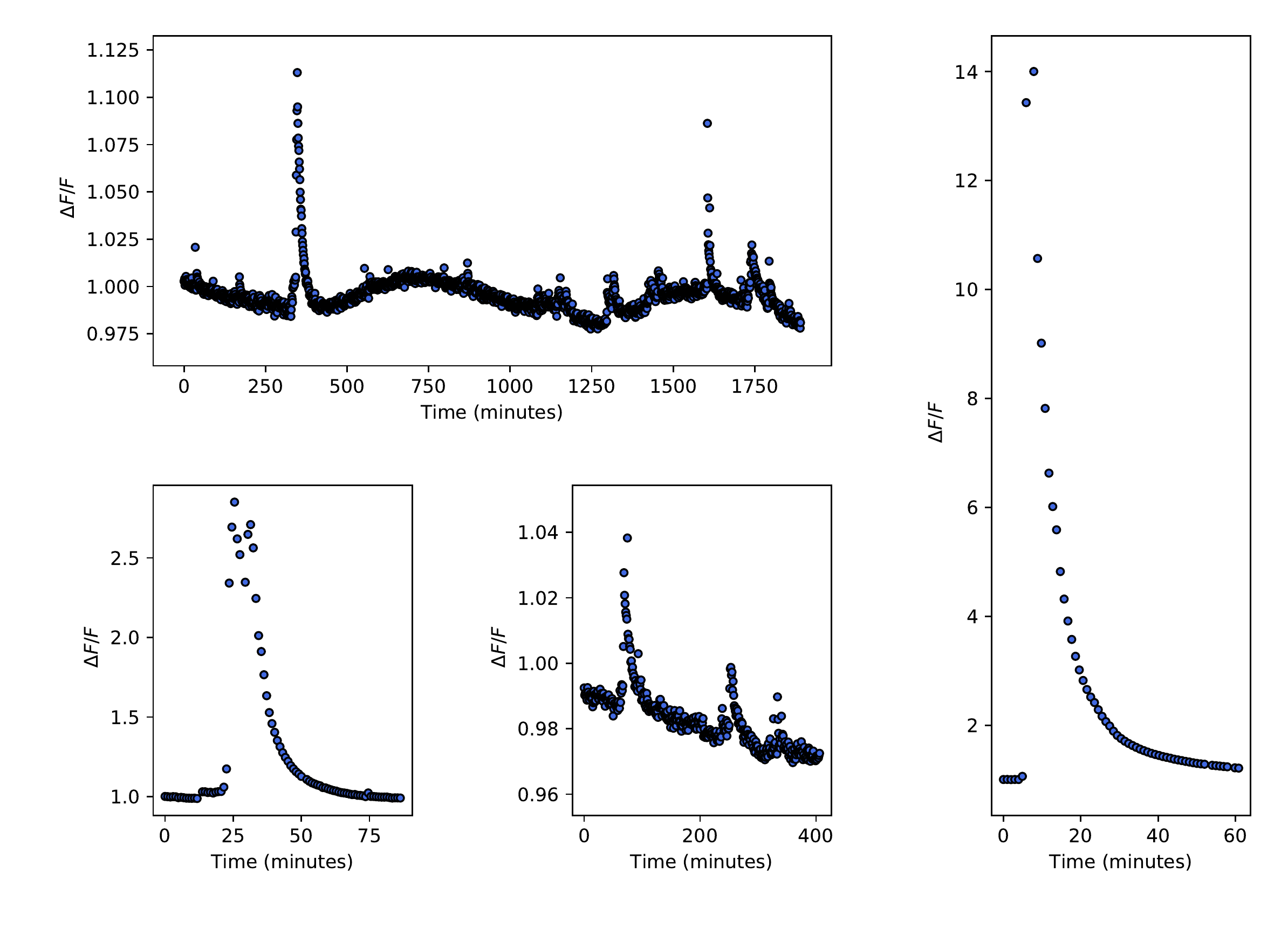}
	\caption{Here we show flares of varying magnitude and duration from the M dwarf DP Cnc (EPIC 211817361) which is an M3.5 spectral type with a rotation period of 0.596 days. In the top right we have a small section of the K2 lightcurve covering $\sim$ 1 day and detailing the frequency of flares on this star. In addition, the left most panel shows the largest flare in the lightcurve of this star normalised peak of 14. The remaining panels show a range of flares with lower energies. }
	\label{flare_example}
\end{figure*}

Our analysis has now proceeded to the M dwarfs with SC data in Campaigns 10 - 18 and consists of $\sim$ 60 M dwarfs. Figure \ref{sample_lightcurve}, shows an example lightcurve from one of the candidates in this extended sample. DP Cnc (EPIC 211817361) has a rotation period of 0.596 days with effective temperature, radius, mass, log g and distance of 3534K, 0.266\Msun, 0.268\Rsun, 5.009 and 38.53pc, respectively (\cite{huber2016vizier}; \cite{gai18}). It is a known flare star according to the {\tt SIMBAD} catalogue and has a spectral type of M3.5. It is an excellent candidate for the further analysis of the distribution of flares with respect to their rotational phase as it is an active star with some very large flares present in its K2 lightcurve, see Figure \ref{flare_example}. In addition, DP Cnc has been observed in Campaigns 16 and 18 which makes it ideal to search for changes in activity. 

\section{Acknowledgments}
We would like to thank Andrew Vanderberg for kindly detrending the K2 short cadence data. We thank the Gaia mission teams and PanStarrs data teams for the respective data in this work. Armagh Observatory and Planetarium is core funded by the Northern Ireland Government through the Dept. for Communities. LD would also like to acknowledge funding from an STFC studentship, a RAS small grant and funding from Cool Stars organisers which made attendance at Cool Stars 20 possible. 

\bibliographystyle{cs20proc}
\bibliography{cool_stars.bib}

\end{document}